\documentclass{ws-mpla}

\def\be{\begin{equation}}
\def\ee{\end{equation}}
\def\bea{\begin{eqnarray}}
\def\eea{\end{eqnarray}}

\begin{document}

\markboth{Robert H. Brandenberger}{String Gas Cosmology and Structure
Formation - A Brief Review}

\catchline{}{}{}{}{}

\title{STRING GAS COSMOLOGY AND STRUCTURE FORMATION - A BRIEF REVIEW
\footnote{Invited talk at CosPA 2006, Nov. 15 - 17,
2006, National Taiwan University, Taipei.}
}

\author{\footnotesize ROBERT H. BRANDENBERGER}

\address{Physics Department, McGill University, 3600 rue Universit\'e,\\
Montr\'eal, QC, H3A 2T8,
Canada\\
rhb@physics.mcgill.ca}

\maketitle


\begin{abstract}

For suitable cosmological backgrounds, thermal fluctuations
of a gas of strings can generate a scale-invariant spectrum
of cosmological fluctuations without requiring a phase of
inflationary expansion. We highlight the key points of
this mechanism, and discuss cosmological backgrounds in which
this scenario can be realized. The spectrum of cosmological
perturbations has a small red tilt (like in scalar field-driven
inflation) but (unlike in inflation) there is a small blue tilt
of the spectrum of gravitational waves.

\end{abstract}

\ccode{PACS Nos.: 98.80.Cq}

\section{Introduction}	

The inflationary scenario \cite{Guth} (see also \cite{Brout,Sato,Starob})
is the current paradigm of early universe
cosmology. In addition to solving some of the key mysteries about
the homogeneous universe which Standard Big Bang (SBB) cosmology does
not address, inflation provides a causal and predictive theory 
(see e.g. \cite{MFB} for a comprehensive review and for
extensive references) for the 
origin of the cosmological fluctuations which are currently
being mapped out both in the distribution of visible matter on
large scales and in the small cosmic microwave background
anisotropies. In spite of these successes, current realizations
of cosmological inflation have serious conceptual problems
\footnote{See e.g. \cite{RHBrev1} for an early analysis
and \cite{RHBrev3,RHBrev5} for more recent discussions of these
problems.} which provide the motivation for looking beyond
inflationary cosmology.

Current inflationary models are based on General Relativity as
the theory of space-time, coupled to scalar fields as the matter
which is responsible for generating the accelerated expansion of
space. In these models, the energy scale at which inflation takes
place is usually of the order $10^{15}$ GeV and thus close to
the string scale and not much lower than the Planck scale. However,
in any approach to quantizing gravity, terms in the effective
action of space-time different from the Einstein-Hilbert term
become important, and thus the applicability of intuition based
on low energy Einstein gravity becomes questionable. This is
one of the key problems facing current realizations of inflation.
A related problem is the {\it cosmological constant problem}. It
is an observational fact that quantum vacuum energy does not
gravitate. Current models of scalar field-driven inflation, however,
use the part of the energy-momentum tensor of scalar field matter
which looks exactly like the contribution of vacuum energy to
generate inflation. Thus, in the absence of knowing what solves the
cosmological constant problem, it is questionable whether it is
legitimate to assume that the constant part of the scalar field
potential energy gravitates. 

Other problems of scalar field-driven inflation include the
{\it singularity problem} - it can be shown that an initial cosmological
singularity is unavoidable \cite{Borde}, like in SBB cosmology -
and the {\it trans-Planckian problem} - the calculation of the
spectrum of cosmological fluctuations is done using an effective
field theory extrapolated beyond its region of validity, and the
results of the calculations are not robust against changes to the
physics in the ultraviolet region (see \cite{Martin} for the
initial study of this problem).

In order to develop a more consistent theory of very early universe
cosmology, new input from fundamental physics is required. Since
superstring theory is the best candidate for a theory which
unifies all forces of nature at very high energies, and thus should
determine the early evolution of the hot primordial universe, we
will review attempts to construct a cosmology of the very early
universe making use of basic ingredients of string theory.

Before starting, we should mention some basic caveats: currently,
we do not have a consistent non-perturbative formulation of string
theory. In order to establish a true theory of string cosmology,
knowledge of non-perturbative string theory will be crucial. Thus,
we are forced at the current stage to focus on toy models of 
string cosmology. The guiding principle of our work (going back
to \cite{BV}) is to focus on the role of new degrees of freedom and
new symmetries of string theory and to study their possible
implications for the evolution of the early universe.

\section{String Theory and a New Cosmological Background}

Let us consider space to be compact, for example a torus of radius $R$.
Closed strings have three types of degrees of freedom. First, there
are the momentum modes which describe the motion of the center of mass
of the string, and whose energies are quantized in units of $1/R$.
In the case of point particle theories, these are the only degrees of
freedom. Closed strings, however, have two other types of degrees of
freedom, namely the oscillatory modes and the winding modes (the latter
label the number of times that the string winds the torus). The energies
of the oscillatory modes is independent of $R$, the winding mode
energies are quantized in units of $R$ (we are setting the string length
to be one to simplify the notation). 

Since the number of oscillatory modes increases exponentially with
energy, there is a limiting temperature, the Hagedorn temperature $T_H$
for a gas of strings in thermal equilibrium \cite{Hagedorn}. This is
a first important consequence of string theory which cannot be seen in
an effective field theory description. The second key feature of
string theory is T-duality, the symmetry of the spectrum of string
states under the interchange of $R$ and $1/R$. As discussed e.g. in
\cite{Pol}, postulating that T-duality is a symmetry of non-perturbative
string theory leads to the existence of D-branes as further fundamental
objects in string theory.

Based on T-duality and the existence of a maximal temperature, we
can already draw some important conclusions for string cosmology
\cite{BV}. Imagine taking a box filled with a thermal bath of strings
and adiabatically reducing the size $R$. At large values of $R$, the
energy will be in the momentum modes which are light. 
Initially, the temperature will increase as in standard cosmology.
However, as $R$ approaches the string scale, thermal equilibrium will
dictate that the energy will drift into the oscillatory modes. The
temperature will approach but never quite reach $T_H$, and the
temperature-radius curve will thus start to deviate from what would be
obtained in standard cosmology. Once $R$ drops below the string scale,
the energy of the string gas will drift into the winding modes which
are now the light states, and the temperature will decrease,
consistent with the symmetry $T(R) = T(1/R)$, as depicted in
Figure 1.

\begin{figure}[th]
\centerline{\psfig{file=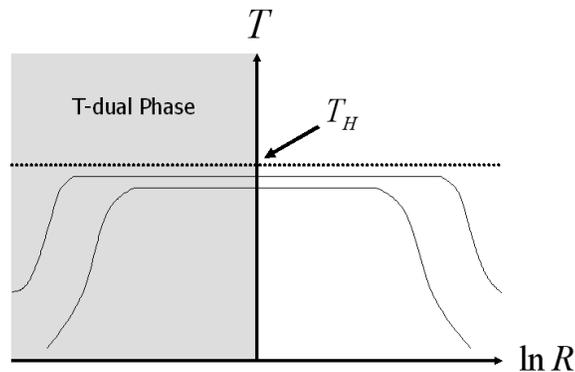,width=3.0in}}
\vspace*{8pt}
\caption{Dependence of the temperature $T$ of a gas of closed
strings (vertical axis) on the radius of the spatial torus $R$
(horizontal axis). The phase when the temperature is to a first
approximation independent of $R$ and close to the Hagedorn
temperature $T_H$ is called the {\it Hagedorn phase}. Its
duration depends on the entropy content of the universe.}
\end{figure}

Let us imagine now that some dynamical principle dictates how
$R$ evolves as a function of cosmic time $t$. There appear to be
two possibilities: if, as we go back in time, $R$ continues to
decrease below the value one, then the temperature-time curve
will evolve as in a non-singular bouncing cosmology, achieving
a temperature maximum
at the point of enhanced symmetry. Alternatively, it is
conceivable that $R$ approaches the enhanced symmetry point only
asymptotically as $t$ decreases, yielding a cosmology similar
to what has been called an {\it emergent universe} \cite{Ellis}.

Note that in the Hagedorn phase, the string gas matter has
vanishing pressure: the positive pressure of the momentum modes
cancels against the negative pressure of the winding modes. As
$R$ expands, the winding modes fall out of equilibrium and
gradually decay (see e.g. \cite{BEK} for a set of equations
describing this process). Thus, a smooth transition from the
Hagedorn phase to the radiation phase of standard cosmology
emerges.

A key lesson which emerges already at this stage of the discussion
is that we should expect that close to the string scale the
usual intuition from General Relativity completely breaks down. We
should expect that at very high densities and at temperatures close
to the Hagedorn temperature, matter satisfying the usual
energy conditions will not
yield a rapid change in temperature. Extrapolating this lesson
slightly, we should also expect that a Hagedorn energy density
does not necessarily lead to rapid expansion of space.

To demonstrate the last point, we need to work in the context of
some toy model which describes the coupling of space-time to
string gas matter. The toy model we choose should be consistent
with the basic symmetry of string theory, namely T-duality.
The simplest action which satisfies this condition is dilaton
gravity, whose action (in the string frame) is given by
\be \label{E0}
S \, = \, - \int d^{1 + N}x \sqrt{-g}e^{-2\phi}
\bigl[R + 4 g^{\mu \nu} \partial_{\mu} \phi \partial_{\nu} \phi \bigr] \, ,
\ee
where $R$ is the string frame Ricci scalar, $g$ is the determinant of
the string frame metric, $N$ is the number of spatial dimensions,
and $\phi$ is the dilaton field. 

The resulting equations of motion for a homogeneous and isotropic
universe with metric
\be \label{metric}
ds^2 \, = \, dt^2 - a(t)^2 d{\bf x}^2 \, ,
\ee
are \cite{TV,Ven}
\bea
-N {\dot \lambda}^2 + {\dot \varphi}^2 \, &=& \, e^{\varphi} E 
\label{E1} \\
{\ddot \lambda} - {\dot \varphi} {\dot \lambda} \, &=& \,
{1 \over 2} e^{\varphi} P \label{E2} \\
{\ddot \varphi} - N {\dot \lambda}^2 \, &=& \, {1 \over 2} e^{\varphi} E \, ,
\label{E3}
\eea
where $E$ and $P$ denote the total energy and pressure, respectively,
and we have introduced the logarithm of the scale factor 
\be
\lambda(t) \, = \, {\rm log} (a(t))
\ee
and the rescaled dilaton
\be
\varphi \, = \, 2 \phi - N \lambda \, .
\ee

Let us make two key observations. First, in the radiation phase of
standard cosmology the driving force for the dilaton vanishes.
Hence, because of the Hubble friction term in its equation of motion,
the dilaton comes to rest, and the two first equations of (\ref{E3})
reduce to the usual Friedmann-Robertson-Walker equations. 
Secondly, in the Hagedorn phase the string gas pressure vanishes.
This implies that the string frame scale factor becomes static.
The dilaton, however, is increasing as we go backwards in time (we take
the branch of solutions of the dynamical equations where this is
true).

Thus, string gas cosmology gives rise to a new cosmological background.
As we follow our presently observed universe back into the past
through the radiation phase of standard cosmology, we smoothly
enter a quasi-static Hagedorn phase, a phase in which the string
frame scale factor is constant. The dilaton is constant in the
radiation phase but begins to increase as we go back in time in
the Hagedorn phase. In fact, the dilaton rapidly blows up
\cite{Betal}. Once the dilaton becomes greater than zero, we are
no longer in the region of validity of dilaton gravity. Instead,
we enter a strongly coupled Hagedorn phase, a good description
of which is still lacking. The basic symmetries of the
physical setup make is reasonable to assume \cite{Betal} that the
dilaton is in fact fixed during the strongly coupled Hagedorn phase.
In this case, there is no difference between the Einstein frame
and the string frame. The strongly coupled Hagedorn phase thus
becomes a static, meta-stable and long-lasting state, allowing
the establishment of thermal equilibrium over distance scales
large compared to the Hubble radius at the beginning of the radiation
phase. In the following section we will see that in the context
of this cosmological background, string theory provides a new
mechanism for producing fluctuations \cite{NBV,Ali,BNPV2}.

Note that the size of our universe in the Hagedorn phase must be
very large (at least $1$ mm if the Hagedorn temperature is of
the order of the scale of Grand Unification - a length obtained by
evolving the physical length of the current Hubble radius back in
time according to the laws of standard cosmology). One of
the successes of inflationary cosmology is that it provides a
mechanism for generating a sufficiently large universe from
a Planck scale primordial universe. String gas cosmology does
not at the present stage of development solve this {\it entropy
problem} without further assumptions. One possibility 
\cite{Natalia} is that an initial phase of nontrivial dynamics
of the bulk spatial dimensions can provide the increase in
entropy \footnote{Note that in the context of a bouncing
cosmology where the universe starts out cold and large, the
entropy problem does not arise.} 

\section{String Gases and Structure Formation}

The new cosmological background discussed in the previous
section yields the space-time diagram sketched in Figure 2.
The vertical axis is time, the horizontal axis denotes the
physical distance. For times $t < t_R$, 
we are in the static Hagedorn phase and the Hubble radius is
infinite. For $t > t_R$, the Einstein frame 
Hubble radius is expanding as in standard cosmology. The time
$t_R$ is when the dilaton starts to decrease. At a slightly
later time, the string winding modes decay, leading to the
transition to the radiation phase of standard cosmology. 

To understand 
why string gas cosmology can lead to a causal mechanism of structure 
formation, we must compare the evolution of the physical 
wavelength corresponding to a fixed comoving scale  
with that of the Einstein frame Hubble radius $H^{-1}(t)$. 
Recall that the Einstein frame Hubble radius separates 
scales on which fluctuations oscillate (wavelengths smaller than 
the Hubble radius) from wavelengths on which the fluctuations are frozen 
in and cannot be affected by micro-physics. Causal micro-physical 
processes can generate fluctuations only on sub-Hubble scales. 
The first 
key point is that for $t < t_i(k)$, the fluctuation mode $k$ is inside
the Hubble radius, and thus a causal generation mechanism for fluctuations
is possible. The second key point is that fluctuations evolve
for a long time during the radiation phase outside the Hubble radius.
This leads to the squeezing of fluctuations which is responsible for
the acoustic oscillations in the angular power spectrum of CMB
anisotropies (see e.g. \cite{BNPV2} for a more detailed discussion
of this point).

\begin{figure}[th]
\centerline{\psfig{file=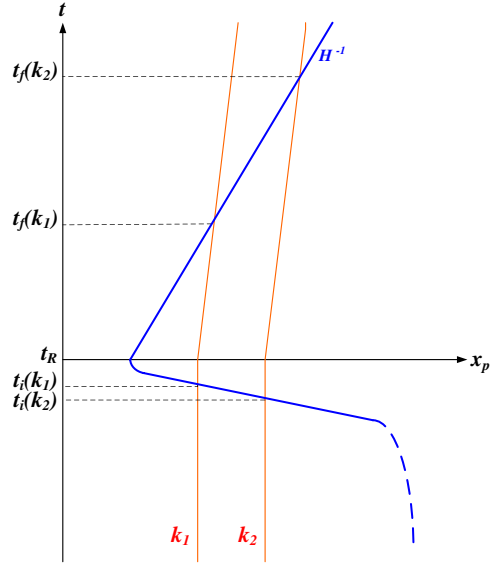,width=3.0in}}
\vspace*{8pt}
\caption{Space-time diagram (sketch) showing the evolution of fixed 
comoving scales in string gas cosmology. The vertical axis is time, 
the horizontal axis is physical distance.  
The solid curve represents the Einstein frame Hubble radius 
$H^{-1}$ which shrinks abruptly to a micro-physical scale at $t_R$ and then 
increases linearly in time for $t > t_R$. Fixed comoving scales (the 
dotted lines labeled by $k_1$ and $k_2$) which are currently probed 
in cosmological observations have wavelengths which are smaller than 
the Hubble radius before $t_R$. They exit the Hubble 
radius at times $t_i(k)$ just prior to $t_R$, and propagate with a 
wavelength larger than the Hubble radius until they reenter the 
Hubble radius at times $t_f(k)$.}
\end{figure}

The corresponding space-time sketch of inflationary cosmology is
given in Figure 3 (the axes are the same as in Figure 2). In
this scenario, the Hubble radius is constant in the phase of
exponential expansion of space, whereas the physical wavelength
of fixed comoving scales expands exponentially. Scales of current
interest in cosmology originate on sub-Hubble scales, are
red-shifted beyond the Hubble radius and are then squeezed for
a long time.

\begin{figure}[th]
\centerline{\psfig{file=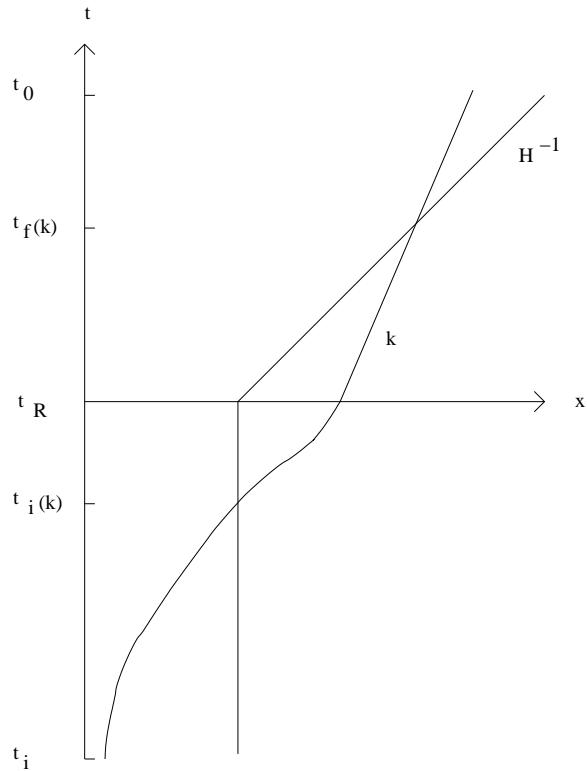,width=3.0in}}
\vspace*{8pt}
\caption{Space-time diagram (sketch) showing the evolution
of scales in inflationary cosmology. The period of inflation 
lasts between $t_i$ and
$t_R$, and is followed by the radiation-dominated phase
of standard big bang cosmology. During exponential inflation,
the Hubble radius $H^{-1}$ is constant in physical spatial coordinates
(the horizontal axis), whereas it increases linearly in time
after $t_R$. The physical length corresponding to a fixed
comoving length scale labelled by its wavenumber $k$ increases
exponentially during inflation but increases less fast than
the Hubble radius after inflation.}
\end{figure}

In both scenarios, fluctuations emerge on sub-Hubble scales, thus
allowing for a causal generation mechanism. However, the mechanisms
are very different. In inflationary cosmology, the exponential
expansion of space leaves behind a matter vacuum, and thus fluctuations
originate as quantum vacuum perturbations. In contrast, in string gas
cosmology space is static at early times, matter is dominated by a
string gas, and it is thus string thermodynamical fluctuations which
seed the observed structures.

The thermodynamics of a gas of strings was worked out some time
ago. We will consider our three spatial dimensions to be compact,
admitting stable winding modes. Specifically, we will take space
to be a three-dimensional torus. In this case, the string gas
specific heat is positive, and string thermodynamics is well-defined,
as was studied in detail in \cite{Deo}. 

Our proposal for string gas structure formation is the following.
For a fixed comoving scale with wavenumber $k$ we compute the matter
fluctuations while the scale in sub-Hubble (and therefore gravitational
effects are sub-dominant). When the scale exits the Hubble radius
at time $t_i(k)$ we use the gravitational constraint equations to
determine the induced metric fluctuations, which are then propagated
to late times using the usual equations of gravitational perturbation
theory (see e.g. \cite{MFB} for an in depth review, and \cite{RHBrev2}
for a pedagogical survey). 

Assuming that the fluctuations are described by the perturbed Einstein
equations (they are {\it not} if the dilaton is not fixed 
\cite{Betal,KKLM}), then the spectra of cosmological perturbations
$\Phi$ and gravitational waves $h$ are given by the energy-momentum fluctuations in the following way \cite{BNPV2}
\be \label{scalarexp} 
\langle|\Phi(k)|^2\rangle \, = \, 16 \pi^2 G^2 
k^{-4} \langle\delta T^0{}_0(k) \delta T^0{}_0(k)\rangle \, , 
\ee 
where the pointed brackets indicate expectation values, and 
\be 
\label{tensorexp} \langle|h(k)|^2\rangle \, = \, 16 \pi^2 G^2 
k^{-4} \langle\delta T^i{}_j(k) \delta T^i{}_j(k)\rangle \,, 
\ee 
where on the right hand side of (\ref{tensorexp}) we mean the 
average over the correlation functions with $i \neq j$. Note that
we have taken the metric including perturbations to be given by
\be \label{pertmetric}
d s^2 = a^2(\eta) \left\{(1 + 2 \Phi)d\eta^2 - [(1 - 
2 \Phi)\delta_{ij} + h_{ij}]d x^i d x^j\right\} \,. 
\ee 
where we have used longitudinal gauge for the cosmological
perturbations. Note that $h$ in (\ref{tensorexp}) indicates the
amplitude of the gravitational waves $h_{ij}$. 

The root mean square energy density fluctuations in a sphere of
radius $R = k^{-1}$ are given by the specific heat
capacity $C_V$ via
\be \label{cor1}
\langle \delta\rho^2 \rangle \,  = \,  \frac{T^2}{R^6} C_V \, . 
\ee 
The result for the specific heat of a gas of closed strings
on a torus of radius $R$ is \cite{Deo}
\be \label{specheat2} 
C_V  \, \approx \, 2 \frac{R^2/\ell_s^3}{T \left(1 - T/T_H\right)}\, , 
\ee 
where $\ell_s$ is the string length.

The power spectrum of scalar metric fluctuations is given by
\bea \label{power2} 
P_{\Phi}(k) \, & \equiv & \, {1 \over {2 \pi^2}} k^3 |\Phi(k)|^2 \\
&=& \, 8 G^2 k^{-1} <|\delta \rho(k)|^2> \, . \nonumber \\
&=& \, 8 G^2 k^2 <(\delta M)^2>_R \nonumber \\ 
               &=& \, 8 G^2 k^{-4} <(\delta \rho)^2>_R \nonumber \\
&=& \, 8 G^2 {T \over {\ell_s^3}} {1 \over {1 - T/T_H}} 
\, , \nonumber 
\eea 
where in the first step we have used (\ref{scalarexp}) to replace the 
expectation value of $|\Phi(k)|^2$ in terms of the correlation function 
of the energy density, and in the second step we have made the 
transition to position space 

The `holographic' scaling $C_V(R) \sim R^2$ is responsible for the
overall scale-invariance of the spectrum of cosmological perturbations. 
In the above equation, for a scale $k$ 
the temperature $T$ is to be evaluated at the
time $t_i(k)$. Thus, the factor $(1 - T/T_H)$ in the 
denominator is responsible 
for giving the spectrum a slight red tilt.

As discovered in \cite{BNPV1}, the spectrum of gravitational
waves is also nearly scale invariant. However, in the expression
for the spectrum of gravitational waves the factor $(1 - T/T_H)$
appears in the numerator, thus leading to a slight blue tilt of
the spectrum. This is a prediction with which the cosmological
effects of string gas cosmology can be distinguished from those
of inflationary cosmology, where quite generically a slight red
tilt for gravitational waves results. The physical reason for the
blue tilt is that
large scales exit the Hubble radius earlier when the pressure
and hence also the off-diagonal spatial components of $T_{\mu \nu}$
are closer to zero.

To summarize this section, we have shown that thermal fluctuations
of a string gas (on a toroidal space) generated during an early
Hagedorn phase of string gas cosmology lead to an almost
scale-invariant spectrum of cosmological fluctuations and gravitational
waves. The spectrum of scalar fluctuations has a slight red tilt
whereas the spectrum of gravitational waves has a small blue tilt,
a distinctive feature of this scenario. As long as the string
energy scale is a couple of orders of magnitude smaller than the
Planck scale, as assumed in early works on heterotic string theory,
the correct amplitude of the fluctuations results.

\section{Towards a Background for String Gas Structure Formation}

Let us review the key requirements for the string gas structure
formation scenario to yield a roughly scale-invariant spectrum.
Firstly, the background cosmology requires a phase in the
early universe during which the Einstein frame Hubble radius rapidly
decreases. At the same time, the cosmological fluctuations need
to obey the generalized Poisson equation on sub-Hubble scales
(which is not the case if the dilaton is rolling as would be
predicted by unmodified dilaton gravity). Finally, the evolution
of fluctuations on super-Hubble scales must be given by the usual
equations of cosmological perturbation theory. 

In the context of the traditional approach to string gas cosmology
\cite{BV,TV}, this requires postulating a strongly coupled Hagedorn
phase during which the dilaton is fixed. There are, however, other
possibilities. One such possibility is described in \cite{Biswas2}.
It is a bouncing cosmology which results from a special higher
derivative gravity action constructed \cite{Biswas} such as to
avoid ghosts. Provided that matter is described by a string gas,
then this gravitational action leads to a Hagedorn phase around
the bounce time. If there was thermal equilibrium in the
contracting phase, then the string gas around the bounce will be
in thermal equilibrium. The dilaton is fixed in this approach,
and therefore matter and metric perturbations are related by the
Poisson equation on sub-Hubble scales. On scales of cosmological
interest today, the effects of the higher derivative terms in
the action are suppressed by powers of $k/M_s$, where $M_s$ is
the characteristic energy scale of the bounce.
Thus, all of the criteria required for our string gas cosmology
mechanism are realized. 

\section*{Acknowledgments}

I wish to thanks the organizers of CosPA06 for inviting me
to lecture, and for their wonderful hospitality. I would like
to thank the staff of the NTU CosPA center for their
marvelous assistance during the conferences. I also would like
to thank all of my collaborators on the string gas cosmology
projects discussed here for the joy of collaboration. In
particular, I thank Sugumi Kanno for giving me permission to
use Figures 1 and 3, originally drawn for \cite{Betal}. My
research is supported in part by an NSERC Discovery grant and 
by funds from the
Canada Research Chair program. Further assistance from a FQRNT
team grant is also acknowledged.


\end{document}